%% file: IMRIPe_Paper.tex
\documentclass[aps,prd,amsmath,floats,floatfix, twocolumn,
superscriptaddress,nofootinbib,showpacs]{revtex4-1}

\input{preamble}

\begin{document}
	
	\title{\TITLE}
	
	\input{authors}

	\date{\today}
	
	\begin{abstract}

Intermediate mass ratio inspiral (IMRI) binaries---containing stellar-mass black holes coalescing into intermediate-mass black holes ($M>100M_{\odot}$)---are a highly anticipated source of gravitational waves (GWs) for Advanced LIGO/Virgo. Their detection and source characterization would provide a unique probe of strong-field gravity and stellar evolution. Due to the asymmetric component masses and the large primary, these systems generically excite subdominant modes while reducing the importance of the dominant quadrupole mode. 
Including higher order harmonics can also result in a $10\%-25\%$ increase in signal-to-noise ratio for IMRIs, which may help to detect these systems. 
We show that by including subdominant GW modes into the analysis we can achieve a precise characterization of IMRI source properties. 
For example, we find that the source properties for IMRIs can be measured to within $2\%-15\%$ accuracy at a fiducial  signal-to-noise ratio of 25 if subdominant modes are included. 
When subdominant modes are neglected, the accuracy degrades to $9\%-44\%$ and significant biases are seen in chirp mass, mass ratio, primary spin and luminosity distances. 
We further demonstrate that including subdominant modes in the waveform model can enable an informative measurement of both individual spin components and improve the source localization by a factor of $\sim$10.
We discuss some important astrophysical implications of high-precision source characterization enabled by subdominant modes such as constraining the mass gap and probing formation channels.
\end{abstract}

\maketitle
	\section{Introduction} 

	Binaries consisting of a stellar-mass black hole and an intermediate-mass black hole (IMBH)---total mass $M\sim\mathcal{O}(10^2\,M_{\odot})$---are a possible source of GWs for the current generation of detectors - Advanced LIGO \cite{TheLIGOScientific:2014jea}, Virgo \cite{TheVirgo:2014hva}, and KAGRA \cite{Akutsu:2020his}.
	Such sources are typically referred to as {\em intermediate mass-ratio inspirals} (IMRIs)\footnote{We note that the merger and ringdown, in addition to the inspiral, are important for detection and source-parameter inference.}. 
	Detection of GWs from IMRIs will shed light on many interesting scientific questions \cite{AmaroSeoane:2007aw, Berry:2019wgg, Sedda:2019uro}: IMRI sources will help us understand the formation channel and evolutionary pathway to supermassive black hole binaries \cite{Bellovary:2019nib}, probe stellar evolution \cite{Gair:2010dx}, and investigate possible environmental effects of matter in the GW signal \cite{AmaroSeoane:2007aw, Barausse:2006vt, Barausse:2007dy, Gair:2010iv, Yunes:2011ws, Barausse:2014tra, Barausse:2014pra, Derdzinski:2020wlw}. 
	IMRI signals could further be used to test general relativity in the strong-field regime \cite{Gair:2012nm, Piovano:2020ooe,Yunes:2009ry, Canizares:2012ji, Canizares:2012is,Rodriguez:2011aa, Chua:2018yng} and to offer an independent measurement of the Hubble constant \cite{MacLeod:2007jd}.
	
	Detectability and parameter estimation accuracy for IMRIs have garnered a lot of interest over the last few years \cite{Mandel:2008bc, Huerta:2010un, Huerta:2010tp, Gair:2010dx, Haster:2015cnn, Leigh:2014oda,MacLeod:2015bpa,Amaro-Seoane:2018gbb}. When formed through hierarchical mergers, IMRI systems will have a large total mass and large spin on the primary. 
	As a motivating example, if a GW190521-like remnant~\cite{Abbott:2020tfl}  ($M \approx 142 M_{\odot}$, $\chi \approx .7$, $z \approx 0.8$)
	captured a $30 M_{\odot}$ stellar-mass BH, the nascent IMRI system would have a detector-frame total mass of around $310 M_{\odot}$. Due to the large total mass characterized by these systems, the number of in-band inspiral cycles from the dominant quadrapole mode is negligible. Previous studies have shown that parameter inference using only the dominant quadrapole mode leads to large uncertainty and significant biases in key source parameters such as the mass and spin of the primary BH~\cite{Haster:2015cnn} and can potentially bias tests of GR \cite{Pang:2018hjb}. This greatly reduces the science that can be extracted from IMRI signals, such as measuring the pair-instability mass-gap \cite{Belczynski:2016jno, Renzo:2020rzx, Farmer:2019jed, Stevenson:2019rcw}, distinguishing between IMRI formation channels \cite{AmaroSeoane:2007aw, Berry:2019wgg}, and self-consistency tests of GR that will be especially informative given the unique IMRI signal \cite{Gair:2012nm, Piovano:2020ooe,Yunes:2009ry, Canizares:2012ji, Canizares:2012is,Rodriguez:2011aa, Chua:2018yng}.

	Fortunately, the asymmetric black hole masses will excite subdominant modes that, due to their higher-frequency content, are in-band longer.
	In this paper, we show that including higher order harmonics into the parameter estimation analysis results in a 3 to 4 times improvement in the measurement uncertainties and 10 times improvement in the recovered 3d comoving volume that contains the true position of the binary.
	We further show that omission of higher-order multipoles leads to either poorer constraints or completely biased estimation of binary properties.  
	We focus on IMRIs with detector-frame total masses $175 M_{\odot} < M < 300 M_{\odot}$ and mass-ratio $1/40<q<1/10$ ($q:=m_2/m_1$ with $m_1 \geq m_2$ and where $m_1$ and $m_2$ are the mass of the primary and secondary black holes respectively). 
	We also demonstrate that high-precision parameter estimates are similarly obtained for generic spin configurations and possible binaries in the pair-instability mass-gap.
	
	The rest of the paper is organized as the follows. Section \ref{Sec:setup} presents a brief outline of the data analysis framework. In Section \ref{Sec:detectability}, we consider the detectability of IMRIs based on a signal-to-noise computation. Parameter estimation results, the main contribution of this paper, is presented in Section \ref{Sec:result}. The robustness of our results is further discussed in Section \ref{Sec:SEOB}. Finally, we discuss the implications, caveats, and conclusions of our analysis in Section \ref{Sec:conlcusion}.
	
	\section{Analysis setup}
	\label{Sec:setup}

	We model the strain data $d$ from a GW detector as a gravitational-wave signal $h$ with an added stream of random noise $n(t)$, often assumed to be Gaussian and stationary \cite{Veitch_2010}, 
	\begin{equation}
	d(t)=h(t; \btheta)+n(t).
	\end{equation}
	The gravitational-wave source parameters $\btheta$ can be inferred from the time-series data using Bayesian inference. Bayesian inference relates the probability of model parameters $\btheta$ to experimental data d, and a hypothesis for the data $\mathcal{H})$, via
		Bayes theorem:
		\begin{equation}
		\label{eq:bayes}
		p(\btheta|d, \mathcal{H}) = \frac{\pi(\btheta| \mathcal{H})\,\mathcal{L}(d|\btheta,\mathcal{H})}{\mathcal{Z}(d|\mathcal{H})}\,.
		\end{equation}
		The quantity $p(\btheta|d, \mathcal{H})$ is the \textit{posterior probability density} of the parameters $\btheta$ given $d$ and $\mathcal{H}$; $\mathcal{L}(d|\btheta,\mathcal{H})$ is the \textit{likelihood} of $d$ given $\btheta$ and $\mathcal{H}$; $\pi(\btheta| \mathcal{H})$ is the \textit{prior} probability of $\btheta$; and $\mathcal{Z}(d|\mathcal{H})$ is the \textit{evidence} (marginalized likelihood) of $d$ given~$\mathcal{H}$. The posterior density is the target for \textit{parameter estimation}, while the evidence is the target for \textit{hypothesis testing}. 
		
		The vector, $\btheta=(\alpha, \delta, \psi,  t_c, d_L, \n, \blambda)$ is a set of 15 parameters that completely characterizes a binary black hole GW signal in general relativity. The vector
		$\blambda:=\{m_1,m_2,\chi_1,\chi_2,\theta_1,\theta_2,\phi_{12},\phi_{jl}\}$ are the intrinsic parameters that describe the binary: the component masses $m_1$ and $m_2$ (with $m_1>m_2$), dimensionless spin magnitudes $\chi_1$ and $\chi_2$, and four angles $\{\theta_1,\theta_2,\phi_{12},\phi_{jl}\}$ describing the spin orientation (cf. Appendix of \cite{Romero-Shaw:2020owr} for definitions of these angles), and $d_L$ is the luminosity distance. The vector $\n := \{\iota, \varphi_c\}$ is the direction of radiation in the source frame: $\iota$ is the inclination angle between the orbital angular momentum of the binary and line-of-sight to the observer, and $\varphi_c$ and $t_c$ are, respectively, the azimuthal angle and time at coalescence. Right ascension $\alpha$ and declination $\delta$ are the sky localization parameters whereas $\psi$ is the polarization angle.
		
	We consider a network of three ground-based detectors: two Advanced LIGO detectors 
	and the Advanced Virgo detector, all operating at their respective design sensitivities \cite{LIGOScientific:2014pky,VIRGO:2014yos} and use a zero noise configuration. Specifically, the synthetic detector data is exactly equal to the expected response due to our GW source. Since detector noise is assumed to be colored Gaussian noise with zero mean, this choice makes our analysis equivalent to an average over an ensemble of analyses which use infinitely many noise realizations~\cite{abbott2017effects}.
	To estimate the PDFs 
	of BBH parameters $p(\btheta | d, H)$, we use the Bayesian inference package \texttt{parallel-bilby}~\footnote{We use \texttt{bilby} 1.0.3 and \texttt{parallel-bilby} 0.1.6.} \cite{ashton2019bilby, smith2019expediting, Romero-Shaw:2020owr} with the \texttt{dynesty} \cite{speagle2020dynesty} sampler. 
	We consider binaries with total masses $175 M_{\odot} \leq M \leq 300 M_{\odot}$ and mass-ratio $1/40 \leq q \leq 1/10$, which would merge in LIGO/Virgo's sensitive band \cite{Smith_2013}. 
	
	\begin{figure}[thb]
		\includegraphics[width=\columnwidth]{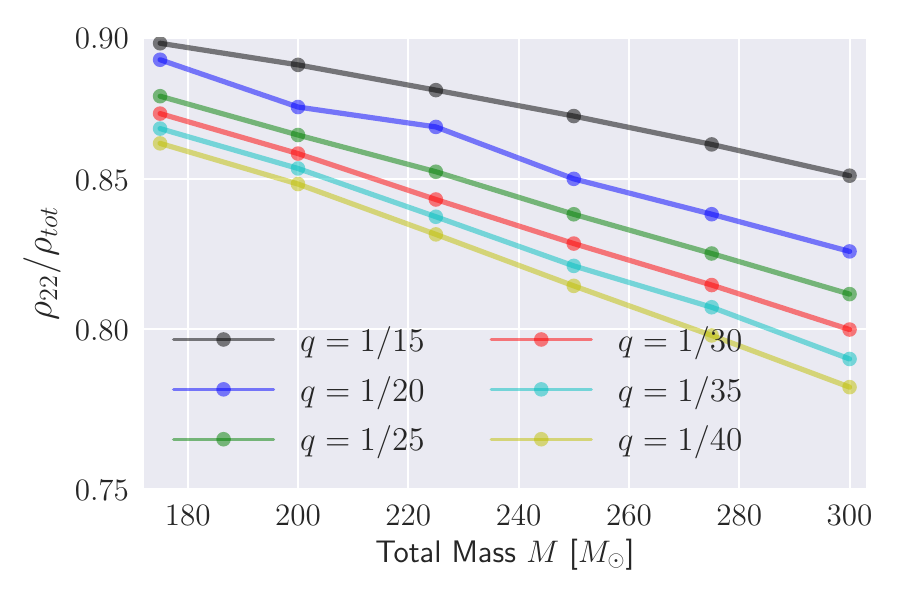}
		\caption{Ratio of the optimal SNRs of only the $(2,\pm2)$ modes, $\rho_{22}$, and all available modes, $\rho_{tot}$, for binaries with total masses $175 M_{\odot}<M<300 M_{\odot}$ and mass-ratio $1/40<q<1/15$. 
			We fix $d_L=750$ Mpc, $\iota=3\pi/4$, $\varphi_c=3\pi/4$, $\alpha=1.0$, $\delta=1.0$, $\psi=0.0$ and $t_c=0.0$h Greenwich mean sidereal time (GMST)(SNR as a function of $\iota$, $\varphi_c$ and $\psi$ is shown in Fig. \ref{Fig:SNR_q30}.) 
		}
		\label{Fig:SNR}
	\end{figure}
	
	\begin{figure}[thb]
		\includegraphics[width=\columnwidth]{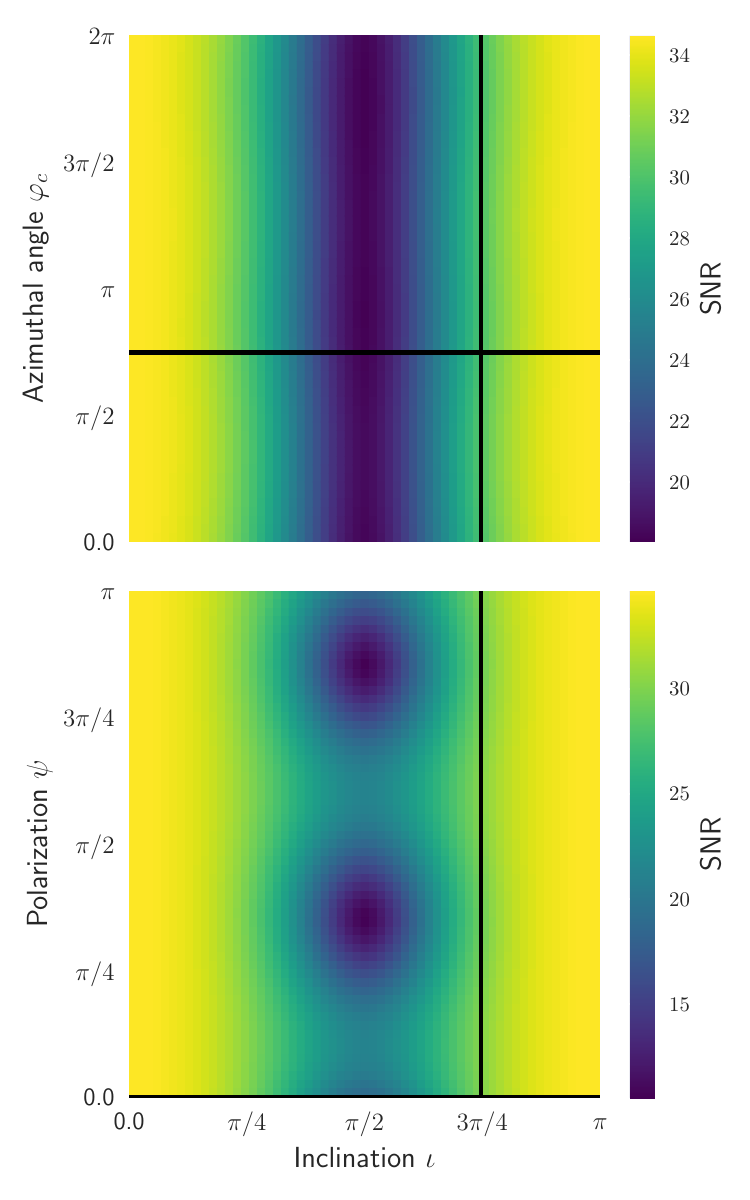}
		\caption{SNR as a function of inclination angle $\iota$, azimuthal angle $\varphi_c$ and polarization $\psi$. We set $q=30$, $M=225 M_{\odot}$, $\{\chi_1,\chi_2\}=\{0.6,0.7\}$, $d_L=750$ Mpc, $\alpha=1.0$ and $\delta=1.0$. Solid black lines indicate our choice of $\iota=2.35$, $\varphi=2.35$ and $\psi=0.0$ for all of our injections.}
		\label{Fig:SNR_q30}
	\end{figure}
	
	We begin by considering binaries whose component spins are either aligned/anti-aligned with the orbital angular momentum. We employ the GW signal model \texttt{IMRPhenomXHM} \cite{Pratten:2020fqn,Garcia-Quiros:2020qpx}~\footnote{The model \texttt{IMRPhenomXHM} has been generated with LALSuite version 6.79, while for analysis using the \texttt{IMRPhenomPXHM} model use LALSuite version 6.83.}., a state-of-art phenomenological non-precessing multi-mode frequency domain model, from the \texttt{LALSuite} software library~\cite{lalsuite}. 
	The model includes $\{\ell,m\}=\{(2,\pm1),(3,\pm3),(3,\pm2),(4,\pm4)\}$ modes 
	in addition to the dominant $\{\ell,m\}=(2,\pm2)$ quadrupolar mode.
	To demonstrate the validity of our results for generic-spin cases, we use \texttt{IMRPhenomXPHM} \cite{Pratten:2020ceb}, a precessing extension of \texttt{IMRPhenomXHM}, that models GW signal emitted by quasi-circular precessing BBHs.

	\begin{figure*}[thb]
		\includegraphics[scale=0.5]{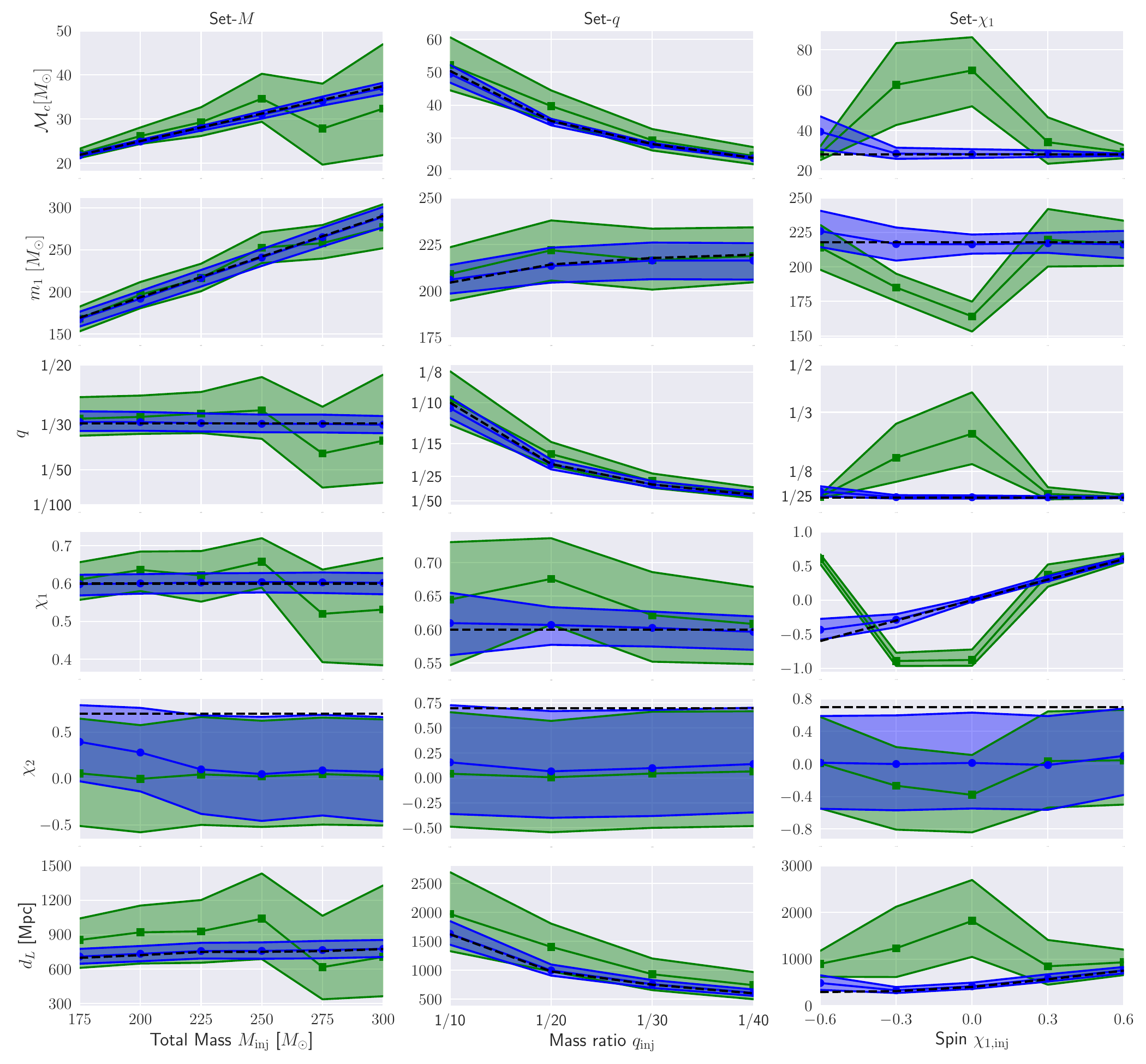}
		\caption{The 90\% credible interval for the total mass of the binary $M$, mass of the primary black hole $m_1$, mass ratio $q$, dimensionless spin parameters $\chi_1$ and $\chi_2$ and luminosity distance $D_L$ as a function of the injected total mass $M_{\rm inj}$ (\textit{left panels}), injected mass ratio $q_{\rm inj}$ (\textit{middle panels}) and injected spin magnitude on the primary black hole $\chi_{\rm 1, inj}$ (\textit{right panels}). 
			90\% credible interval for the $(2,\pm2)$ mode (all modes) recovery is shown in green (blue) and 
			the true values are plotted as a dashed black line. Columns 1, 2, and 3 correspond to BBH configurations of Set-$M$, Set-$q$, and Set-$\chi_1$, respectively.
		}
		\label{Fig:90CI}
	\end{figure*}
		
	We choose uniform priors for chirp masses ($5 M_\odot< \mathcal{M}_{c} < 80 M_\odot$) and mass ratio ($1/150 < q < 1$), defined in the detector frame. 
	Unless otherwise specified, mass parameters in this paper are always reported in the detector frame. This particular choice is made as the detector frame masses are the directly observed quantities whereas source-frame masses are inferred using the estimated luminosity distance $d_L$. This introduces additional uncertainties in the inferred source-frame masses.
	For the component dimensionless spins, we use aligned-spin priors \cite{Lange:2018pyp}. 
	The prior on the luminosity distance is taken to be: $P(d_L) \propto d_L^2$, with $20 \leq D_L \leq 3000$ Mpc. 
	For the orbital inclination angle $\iota$, we assume a uniform prior over $-1 \leq \cos\iota \leq 1$. 
	Priors on the sky location parameters $\alpha$ and $\delta$ are assumed to be uniform over the sky with periodic boundary conditions.

	\section{Detectability of high mass-ratio massive binaries}
	\label{Sec:detectability}
	
	To assess the detectability of the IMRIs in current-generation detectors \cite{Mandel:2008bc, Smith:2013mfa}, we generate signals using \texttt{IMRPhenomXHM} and compute the optimal network signal-to-noise ratio (SNR)~\cite{Sathyaprakash:2009xs} $\rho$ at different points of the parameter space.
	In general, a network SNR of 8-10 is often sufficient for the detection of a GW signal. However, the statistical significance of a detection at a given SNR is established empirically \cite{2020arXiv201014527A}. 
	In Fig.\ref{Fig:SNR}, we show the ratio of the optimal SNR of only the dominant $(\ell,m)=(2,\pm2)$ mode and including the higher order modes as a function of total mass of the binary $M$ for different mass-ratios $q$. 
	We find a significant SNR loss (10\% to 25\%) when disregarding the higher-order modes, where a $(\ell,m)=(2,\pm2)$-only analysis would be likely to miss otherwise clear detections of IMRIs.
	If the detector noise were truly Gaussian, this SNR loss corresponds to a reduction in the volume of the Universe (which is roughly $\sim\rho^3$) from which clear detections can be made of the order of $30 - 60 \%$.
	Due to non-Gaussian detector glitches, however, the sensitivity is likely to be degraded. Therefore, IMRIs are at present difficult to detect with template-based searches~\cite{dal2013effect}.
	We further note that the SNR in both the dominant $(\ell,m)=(2,\pm2)$ mode and in higher order modes are significantly greater than the detection threshold of $\sim 8-10$ for the binaries we consider in this work and so may be detected with other methods~\cite{bustillo2018sensitivity,chandra2020nuria}.
	The inclusion of higher-order modes in modelled GW search pipelines is outside the scope of this paper but would warrant further explorations (see also~\cite{Harry:2017weg} for effects on a more equal-mass BBH search).

	\paragraph{SNR variation with source orientation: }
	For all of our simulated events, we fix the inclination angle $\iota$=2.35, azimuthal angle $\varphi$=2.35 and polarization angle $\psi$=0.0. Varying these values can raise or lower the SNR as the different harmonic modes can undergo constructive or destructive interference. In Figure \ref{Fig:SNR_q30}, we pick up a representative case of mass ratio $q=30$ and show the SNR as a function of $\iota$, $\varphi$ and $\psi$ while keeping other parameters fixed. We set the total mass $M=225 M_{\odot}$, spins $\{\chi_1,\chi_2\}=\{0.6,0.7\}$, luminosity distance $d_L=750$ Mpc, right acession $\alpha=1.0$ and declination $\delta=1.0$. Our choice of $\iota(=2.35)$, $\varphi(=2.35)$ and $\psi(=0.0)$ (black lines) for the simulated events fits in between optimal and conservative SNR expectations.

	\section{Parameter estimation results}
	\label{Sec:result}
	
	We now investigate the precision with which current-generation detectors will be able to measure the source properties of the IMRIs. We study the effects of higher order modes in parameter estimation and explore the validity of our results for binaries with generic spins, as well as special cases, e.g. face-on/off binaries.
	\paragraph{Simulated GW signals: }
	We employ \texttt{IMRPhenomXHM} to simulate the following sets of aligned-spin IMRI signals:
	\begin{itemize}
		\item \textbf{Set-$M$:} We fix $q=1/30$, $\chi_1=0.6$, and $\chi_2=0.7$ while varying the
		total mass $M=\{175,200,225,250,275,300\} M_{\odot}$.
		This set of spins matches with the predicted spin magnitudes of black-holes formed through hierarchical mergers \cite{Gerosa:2017kvu, Fishbach:2017dwv, Berti:2008af}. 
		\item \textbf{Set-$q$:}  We fix $M=225 M_{\odot}$, $\chi_1=0.6$, and $\chi_2=0.7$ while varying
		the mass ratio $q=\{1/40,1/30,1/20,1/10\}$.
		\item \textbf{Set-$\chi_1$:} We fix $M=225 M_{\odot}$, $\chi_2=0.7$, and $q=1/30$
		while varying the dimensionless spin magnitude on the primary black hole $\chi_1=\{-0.6,-0.3,0.0,0.3,0.6\}$.
		\item \textbf{Set-$\chi_2$:} We fix $M=225 M_{\odot}$, $q=1/30$, and $\chi_1=0.6$ while varying
		the dimensionless spin magnitude on the secondary black hole $\chi_2=\{-0.7,-0.3,0.0,0.3,0.7\}$.
	\end{itemize}
	For each binary, we set $\iota=3\pi/4$, $\varphi_c=3\pi/4$, $\alpha=1.0$, $\delta=1.0$ and $\psi=0.0$ radians respectively. 
	We further fix $t_c=0.0$h Greenwich mean sidereal time (GMST).
	The distance to each source is then scaled so that the network SNR is $\rho=25$ to ensure a fair comparison of PE accuracy across the parameter space. 
	We use the same waveform model for injection and recovery thereby removing possible biases due to the choice of a particular waveform approximant~\cite{Smith:2013mfa, Haster:2015cnn, Purrer:2019jcp}.
	We demonstrate the robustness of our results with respect to waveform model in Sec.~\ref{Sec:SEOB}.
	For each simulated IMRI, we recover the source parameters with two different mode configurations: one with only the dominant $(\ell,m)=(2,\pm2)$ modes and then with all available modes included. 
	The injected signal, however, always contains all available modes. 
	
	\begin{table*}
		\caption{Parameter estimation accuracy for the binaries used in our analysis. 
			\footnote{
			The two missing entries for Set-$\chi_{1}$ corresponds to the case where injected value of $\chi_{1}=0.0$ because of which $\Delta \theta/\theta_{\rm inj}$ is undefined.
			}
			We report the dimensionless uncertainties $\Delta \theta/\theta_{\rm inj}$ and biases $\beta_{\theta}$  for
			five representative parameters.
			Symbols: $\mathcal{M}_{c}$: Detector frame chirp mass; $q$: mass ratio; $\chi_1$: dimensionless spin on the primary black hole; $V_{22}$ and $V_{hm}$ are the recovered 90\% credible region of the comoving volume (computed using $\texttt{ligo-skymap}$) containing the true position of the binary with and without the higher modes respectively. Bias in parameter $\lambda$ computed as $\beta_{\theta} = |\theta_{\rm true} -\theta_{\rm recovered}|$ where $\theta_{\rm true}$ is the true value of the parameter and $\theta_{\rm recovered}$ is the median of the recovered posterior for $\theta$. 
			Values in parenthesis denote uncertainties and biases when only $(2,\pm2)$ mode is used in the recovery model.
		}
			\begin{ruledtabular}
					
			\begin{tabular}{l | c c c c c | c c c c c | c}		
					
				Binaries &&& $\frac{\Delta \theta}{\theta_{\rm inj}}$ (\%)&& &&&Biases&& & \\ 
				\hline
				$\{\mathcal{M}_c,\;\;m_1,\;1/q,\;\chi_1,\;\chi_2,\;d_L\}$&$\frac{\Delta \mathcal{M}_c}{\mathcal{M}_{c,\rm inj}}$&$\frac{\Delta m_1}{m_{1,\rm inj}}$&$\frac{\Delta 1/q}{1/q_{\rm inj}}$&$\frac{\Delta \chi_1}{\chi_{1,\rm inj}}$ &$\frac{\Delta d_L}{d_{L,\rm inj}}$ &$\beta_{\mathcal{M}_c}$ &$\beta_{m_1}$ &$\beta_{1/q}$ &$\beta_{\chi_1} $ &$\beta_{d_L}$ &$\frac{V_{22}}{V_{hm}}$\\
				\hline
				\textbf{Set}-$M$ &&&&& &&&&& &\\
				$\{21.86,169.35,30,+0.6,+0.7,697.84\}$ &2.3 &10.3 &16.9 &9.0 &18.6 &0.077 &1.540 &0.32 &0.0015 &10.99 &9.32	\\
				&(9.6) &(17.1) &(33.1) &(16.5) &(61.7)    &(0.326) &(1.13) &(1.13) &(0.0108) &(155.6)  & \\
				$\{24.98,193.54,30,+0.6,+0.7,720.69\}$ &3.0 &9.5 &16.2 &8.5 &18.5 &0.086 &1.601 &0.28 &0.0004 &12.43 &11.12	\\
				&(14.9) &(16.1) &(32.9) &(17.4) &(70.1)    &(1.169) &(3.087) &(1.54) &(0.0359) &(199.4) & \\
				$\{28.11,217.74,30,+0.6,+0.7,749.13\}$ &3.8 &9.0 &15.5 &8.6 &18.2 &0.1641 &1.467 &0.02 &0.0026 &7.89 &13.06\\
				&(23.1) &(14.9) &(35.3) &(22.3) &(72.9)    &(1.167) &(1.254) &(2.31) &(0.0214) &(179.8) & \\
				$\{31.23,241.93,30,+0.6,+0.7,750.10\}$ &5.3 &8.3 &15.2 &8.5 &18.9 &0.236 &1.167 &0.15 &0.0036 &8.35 &24.80\\
				&(34.8) &(15.1) &(53.1) &(21.7) &(99.4)    &(3.386) &(10.591) &(3.00) &(0.0578) &(288.6) & \\
				$\{34.35,266.13,30,+0.6,+0.7,756.02\}$ &5.9 &8.4 &15.3 &9.0 &20.0 &0.282 &1.067 &0.17 &0.0031 &9.40 &7.36\\
				&(53.4) &(15.0) &(69.4) &(40.7) &(96.3)    &(6.524) &(8.363) &(10.53) &(0.0800) &(139.4) & \\
				$\{37.47,290.32,30,+0.6,+0.7,774.01\}$ &6.9 &8.3 &14.7 &9.3 &19.1 &0.454 &1.327 &0.36 &0.0020 &1.99 &13.81\\
				&(67.2) &(18.0) &(92.8) &(47.2) &(124.8)    &(5.108) &(13.619) &(5.31) &(0.0686) &(68.2) & \\

				\hline 
				\textbf{Set}-$q$ &&&&& &&&&& &\\
				$\{28.11,204.5,10,+0.6,+0.7,1620.1\}$ &10.6 &7.4 &17.1 &15.5 &25.2 &0.894 &1.54 &0.46 &0.0097 &8.31 &9.29\\
				&(14.1) &(43.5) &(30.6) &(84.2) &(11.3) &(1.836) &(4.59) &(0.23) &(0.044) &(351.8) & \\
				$\{28.11,214,2,20,+0.6,+0.7,978.16\}$ &5.6 &8.7 &15.8 &9.3 &19.6 &0.287 &0.80 &0.13 &0.0069 &17.95 &15.88\\
				&(26.4) &(13.5) &(39.1) &(21.5) &(84.3)    &(4.529) &(4.93) &(2.78) &(0.075) &(424.2) & \\
				$\{28.11,217.7,30,+0.6,+0.7,749.13\}$ &3.8 &9.0 &15.5 &8.6 &18.2 &0.164 &1.46 &0.03 &0.0026 &7.89 &13.06\\
				&(23.1) &(14.9) &(35.3) &(22.3) &(72.9)    &(1.167) &(1.25) &(2.31) &(0.021) &(179.8) & \\
				$\{28.11,219.5,40,+0.6,+0.7,598.04\}$ &2.8 &8.9 &15.2 &8.3 &18.5 &0.101 &3.24 &0.66 &0.0032 &11.15 &13.10\\
				&(21.8) &(13.4) &(35.9) &(19.2) &(78.5) &(0.631) &(0.682) &(1.84) &(0.008) &(141.8) & \\
				\hline 
				
				\textbf{Set}-$\chi_1$  &&&&& &&&&& &\\
				$\{28.11,217.7,30,-0.6,+0.7,292.01\}$ &58.7 &12.1 &106.2 &50.6 &107.1 &11.23 &8.18 &11.63 &0.165 &192.59 &9.41\\
				&(24.4) &(14.9) &(37.1) &(24.1) &(190.0) &(0.40) &(3.73) &(1.65) &(1.20) &(604.3) & \\
				$\{28.11,217.7,30,-0.3,+0.7,309.83\}$ &19.7 &11.0 &36.6 &64.1 &40.9 &0.39 &1.07 &0.9 &0.012 &17.72 &215.33\\
				&(144.7) &(9.1) &(613.7) &(64.7) &(487.2)    &(34.50) &(32.79) &(24.25) &(0.59) &(920.6) & \\
				$\{28.11,217.7,30,0.0,+0.7,400.15\}$ &15.4 &6.3 &33.7 &- &33.3 &0.11 &1.39 &0.57 &0.003 &14.75 &306.86\\
				&(122.1) &(9.8) &(758.7) &- &(412.2)    &(41.67) &(53.50) &(26.15) &(0.87) &(1416.8) & \\
				$\{28.11,217.7,30,+0.3,+0.7,564,22\}$ &11.3 &6.7 &21.6 &25.8 &27.6 &0.07 &0.66 &0.36 &0.011 &18.91 &22.26\\
				&(81.9) &(19.1) &(127.6) &(108.6) &(170.1)    &(6.04) &(1.89) &(8.04) &(0.08) &(279.1) & \\
				$\{28.11,217.7,30,+0.6,+0.7,759.13\}$ &3.8 &9.0 &15.5 &8.6 &18.2 &0.16 &1.46 &0.03 &0.002 &7.89 &13.06\\
				&(23.1) &(14.9) &(35.3) &(22.3) &(72.9)    &(1.16) &(1.25) &(2.31) &(0.02) &(179.8) & \\
				\hline
				
				\textbf{Set}-$\chi_2$  &&&&& &&&&& &\\
				$\{28.11,217.7,30,+0.6,-0.7,736.13\}$ &4.19 &9.6 &16.5 &9.6 &19.2 &0.16 &1.46 &0.75 &0.0093 &18.32 &10.97\\
				&(27.0) &(15.5) &(40.4) &(25.9) &(80.4) &(0.94) &(5.94) &(2.91) &(0.005) &(182.5) & \\
				$\{28.11,217.7,30,+0.6,-0.3,742.08,\}$ &4.18 &9.6 &16.2 &9.6 &18.3 &0.08 &1.38 &0.52 &0.0056 &14.47 &14.02\\
				&(25.4) &(15.2) &(38.2) &(25.4) &(77.0)    &(0.54) &(5.71) &(2.23) &(0.006) &(157.8) & \\
				$\{28.11,217.7,30,+0.6,0.0,746.52,\}$ &4.24 &9.7 &16.1 &9.7 &18.8 &0.003 &1.08 &0.28 &0.0021 &11.80 &11.41\\
				&(24.3) &(15.0) &(37.8) &(24.2) &(74.3)    &(0.12) &(5.16) &(1.48) &(0.007) &(131.3) & \\
				$\{28.11,217.7,30,+0.6,+0.3,751.19\}$ &4.33 &9.5 &16.1 &9.3 &19.1 &0.06 &1.14 &0.16 &0.0001 &10.77 &10.09\\
				&(24.8) &(15.1) &(36.9) &(24.1) &(76.5)    &(0.38) &(3.87) &(1.58) &(0.002) &(139.7) & \\
				$\{28.11,217.7,30,+0.6,+0.7,759.13\}$ &3.8 &9.0 &15.5 &8.6 &18.2 &0.16 &1.46 &0.03 &0.002 &7.89 &13.06\\
				&(23.1) &(14.9) &(35.3) &(22.3) &(72.9)    &(1.16) &(1.25) &(2.31) &(0.02) &(179.8) & \\
			\end{tabular}
		\end{ruledtabular}
		\label{tab:1}
	\end{table*}
		
	\paragraph{Parameter estimation accuracy: }
	In Fig.\ref{Fig:90CI}, we show the recovered 90\% credible intervals for five important binary source properties: chirp mass $\mathcal{M}_{c}$, mass of the primary black hole $m_1$, mass ratio $q$, spin magnitudes $\chi_1$ and luminosity distance $d_L$ as a function of the injected total mass $M_{\rm inj}$, mass ratio $q_{\rm inj}$ and primary spin $\chi_{1, \rm inj}$.
	Similar to~\cite{Haster:2015cnn}, we find that the best constrained parameters are chirp mass $\mathcal{M}_{c}$, mass ratio $q$ and spin on the primary black hole $\chi_1$. 
	The measurement accuracy of $\mathcal{M}_{c}$ depends on the number of in-band inspiral cycles. 
	As the total mass of the binary increases, the observable signal becomes dominated by the merger and ringdown part. 
	Therefore, the uncertainty on $\mathcal{M}_{c}$ is expected to increase with increasing $M$. 
	We find that $\mathcal{M}_{c}$ can be measured with an accuracy of $\sim3\%$ for a binary with total mass $M=175M_{\odot}$ (and $q=1/30$) while the uncertainty increases to $\sim7\%$ for $M=300M_{\odot}$ (and $q=1/30$). 
	In Table \ref{tab:1}, we summarize the uncertainties on $\mathcal{M}_{c}$, $m_1$, $q$, $\chi_1$ and $d_L$ along with the biases in estimation for binaries at the boundary of our parameter space.
	Since we are using zero noise, our results are equivalent to ensemble averages. This implies that the bias parameter $\beta_{\theta} = \theta_{\rm true} -\theta_{\rm recovered}$ (where $\theta_{\rm true}$ is the true value of the parameter and $\theta_{\rm recovered}$ is the median of the recovered posterior for $\theta$) is, in some sense, the exact quantification of bias from the injected value.
	For typical systems, $m_1$, $q$ and $\chi_1$ is well constrained with $\sim10\%$ of accuracy when higher modes are included. 
	Relative errors on $M$ ($\chi_{\rm eff}$) closely follows that of $m_1$ ($\chi_1$) (cf. Figures \ref{90CI_diffM_supple} and \ref{90CI_diffq_supple}). As the binary becomes more asymmetric, number of waveform cycles in the detector band increases resulting a decrease in measurement uncertainties of $q$.
	Uncertainties in $d_L$ are typically $\sim20$\%.
	When all other parameters are fixed, a negative spin on the primary black hole reduces the number of in-band cycles in gravitational waveform implying a severe loss of information in the detected signal. 
	This leads to significant increase in uncertainties on almost all the recovered parameters for the binary with $\{M,q,\chi_1,\chi_2\}=\{225M_{\odot},1/30,-0.6,+0.7\}$.

	\begin{figure}[thb]
		\includegraphics[width=\columnwidth]{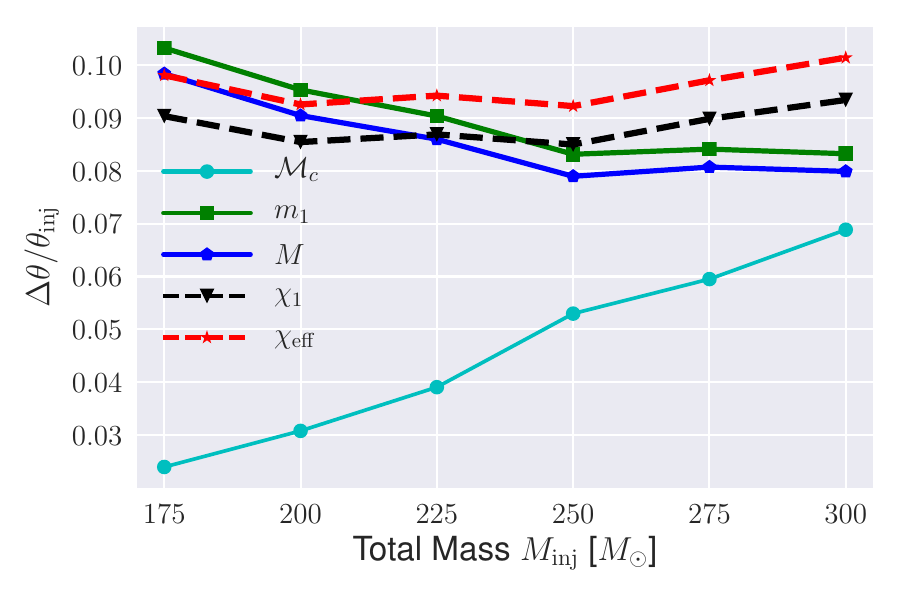}
		\caption{Dimensionless 90\% uncertainties $\Delta \theta/\theta_{\rm inj} (\%)$ for chirp mass $\mathcal{M}_c$, total mass $M$, mass of the primary black hole $m_1$, spin on the primary black hole $\chi_1$, and effective inspiral spin $\chi_{\rm eff}$ as a function of the injected total mass (corresponding to BBH configurations Set-$M$).}
		\label{Fig:90CI_diffM}
	\end{figure}
	
	\begin{figure}[thb]
		\includegraphics[width=\columnwidth]{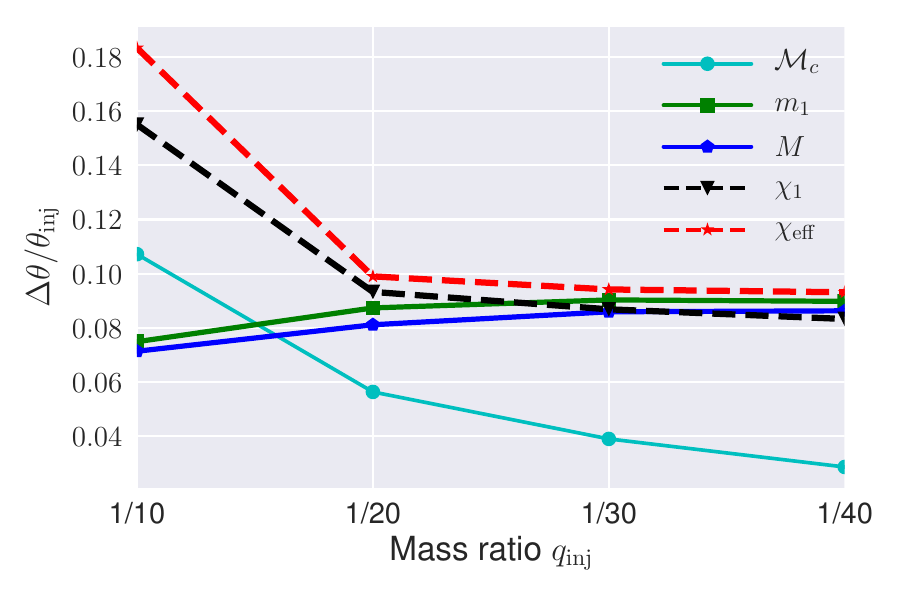}
		\caption{Dimensionless 90\% uncertainties $\Delta \theta/\theta_{\rm inj}  (\%)$ for chirp mass $\mathcal{M}_c$, total mass $M$, mass of the primary black hole $m_1$, spin on the primary black hole $\chi_1$, and effective inspiral spin $\chi_{\rm eff}$ as a function of the injected mass ratio (corresponding to BBH configurations Set-$q$).}
		\label{Fig:90CI_diffq}
	\end{figure}
	
	\begin{figure}[b]
		\includegraphics[width=\columnwidth]{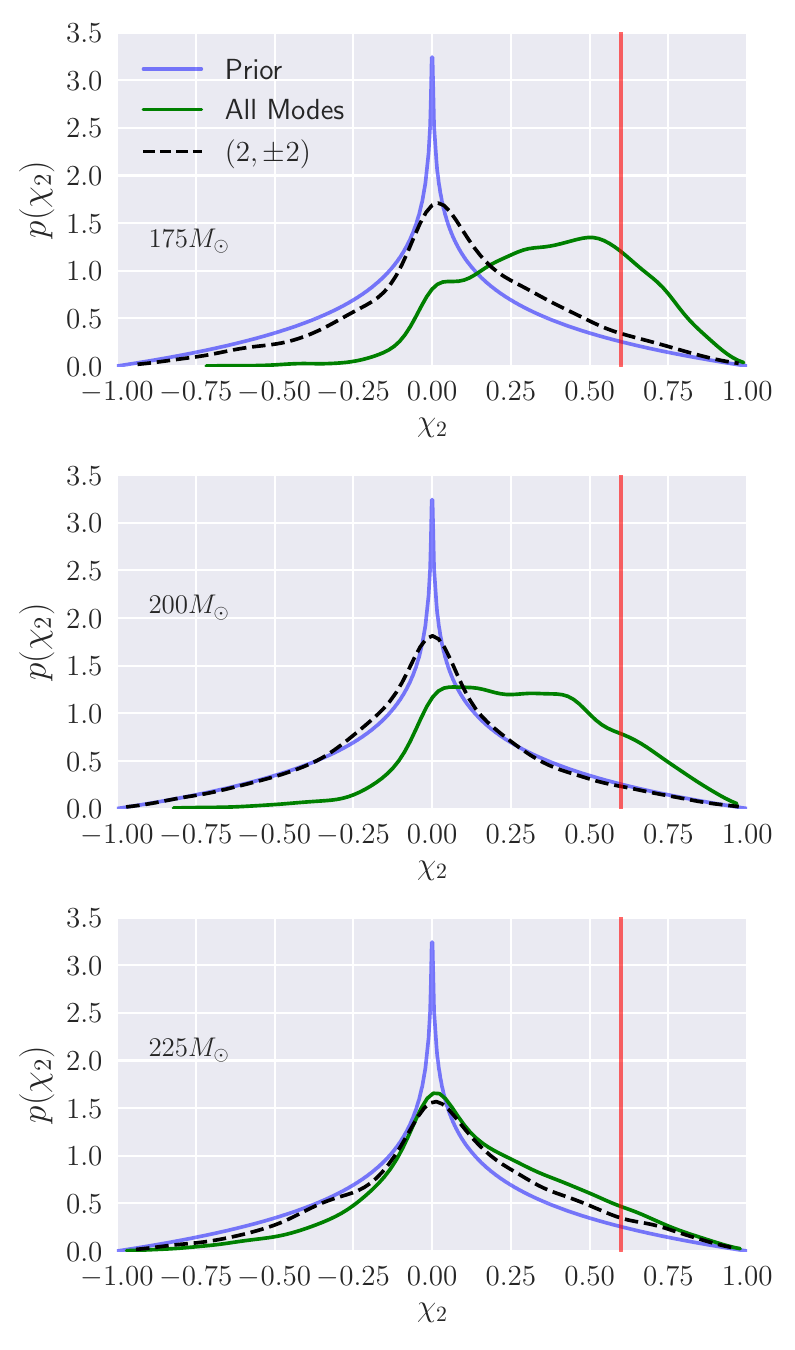}
		\caption{Posterior for the secondary spin parameter $\chi_2$ recovered with (solid green) and without the higher modes (dashed black) for binaries with varying total masses (\textbf{Set-$M$} injections. Blue solid lines show aligned-spin prior \cite{Lange:2018pyp}). 
			All other parameters are set to the default values described in the text. 
			Vertical red lines indicate the true value.}
		\label{Fig:chi2}
	\end{figure}
	
	\paragraph{Constraints on total mass $M$ and effective spin $\chi_{\rm eff}$ :}
	We compute the dimensionless 90\% credible uncertainties for the total mass of the binary, $M$, 
	and the effective inspiral spin, $\chi_{\rm eff}$. We find that, due to the smaller black hole being well approximated by structureless point particle,
	the relative errors on $M$ ($\chi_{\rm eff}$) closely follows that of $m_1$ ($\chi_1$). In Fig. \ref{Fig:90CI_diffM} and \ref{Fig:90CI_diffq}, we show the 90\% uncertainties for different parameters as a function of the injected total mass (BBH configurations Set-$M$) and mass ratio (BBH configurations Set-$q$).
	
	\paragraph{Measureability of Spin Magnitudes: }
	We now consider individual spin measurements for IMRIs, using an aligned-spin prior (solid blue lines) for the both spins.
	While $\chi_1$ can be precisely measured for most of the binaries (as seen in Fig.\ref{Fig:90CI}), the spin on the secondary black hole $\chi_2$ generally remains uninformative. 
	However, contrary to our general expectation, we find that one may be able to constraint $\chi_2$ due to the presence of higher modes. 
	We find that the 90\% credible intervals for $\chi_2$ may be reasonably resolved for our binaries with masses $M<225M_{\odot}$. 
	For larger total masses, the number of resolvable inspiral-cycles reduces drastically implying a sharp drop in available information in the detected signal. 
	In Fig.~\ref{Fig:chi2}, we show the $\chi_2$ posteriors recovered with only $(2,\pm2)$ modes (black dashed lines) and with all modes (green solid lines) for the injections created in $\textbf{Set-$M$}$ with varying total masses. 
	It shows that while the $(2,\pm2)$ mode recovery cannot constrain $\chi_2$, this parameter can be measured when the recovery model includes higher order modes and the signal contains a sufficient number of resolvable inspiral cycles.

	\paragraph{Importance of higher-order modes: }
	The impact of higher modes in detection and parameter estimation has been extensively studied 
	in the comparable- to moderate-mass-ratio regime ($q \lesssim 8$) often using Fisher matrix-based studies or reanalyzing novel gravitational-wave events~\cite{Varma:2016dnf, bustillo2016impact, bustillo2017detectability, Capano:2013raa, Littenberg:2012uj, Bustillo:2016gid, Brown:2012nn, Varma:2014, Graff:2015bba, Harry:2017weg, Kalaghatgi:2019log, kumar2019constraining, islam2020improved, shaik2020impact}.
	Our fully Bayesian results, which focus on plausible IMRI systems for the upcoming LVK observing run, are in broad agreement with these previous works: 
	higher-order modes are increasingly important as the value of the mass ratio increases and/or effective spin decreases, and recovery models that include all modes significantly reduce bias in all cases. 
	As compared to previous IMRI studies using only the quadrapole mode~\cite{Haster:2015cnn}, however, the extent to which higher-order modes enable precise measurements of most system parameters is surprising. 
	This is due to the unique ability of massive IMRIs to excite sufficiently loud higher-order modes that lie in the detector's sensitive band.

	To probe the impact of these subdominant multipoles within our analysis setup, we recover the injected signal with (i) only $(2,\pm2)$ modes and (ii) with all available modes including the dominant $(2,\pm2)$ mode.
	We observe that (i) the 90\% credible interval becomes significantly tighter when higher modes are included in recovery model and (ii) omission of higher modes results in substantial bias in parameter estimation for most binaries (cf. Table~\ref{tab:1} and Figure~\ref{Fig:90CI}).
	For signals with varied primary spin (\textbf{Set-$\chi_{1}$}), we find significant bias whenever $\chi_{1, \rm inj} < 0.3 $ with increasing bias as $\chi_{1, \rm inj}$ is lowered.
	While this effect is well-known from comparable mass-ratio studies with negative spin~\cite{shaik2020impact,bustillo2016impact,Varma:2016dnf} (cf. Fig.~11 of Ref.~\cite{shaik2020impact}), what is particularly striking is that for massive IMRIs noticeable bias occurs even for positive spins as large as $\chi_{1, \rm inj} \approx 0.3$.
	We further find that, when higher modes are included, 90\% credible region of the recovered co-moving volume that contains the true position of the binary is shrunk by almost 10 times (Table \ref{tab:1}). 
	Taken together, these results demonstrate that higher modes will play an especially central role in analyzing signals from high mass ratio massive binaries, including source localization and precise estimation of source properties.

	\begin{figure}[thb]
		\includegraphics[width=\columnwidth]{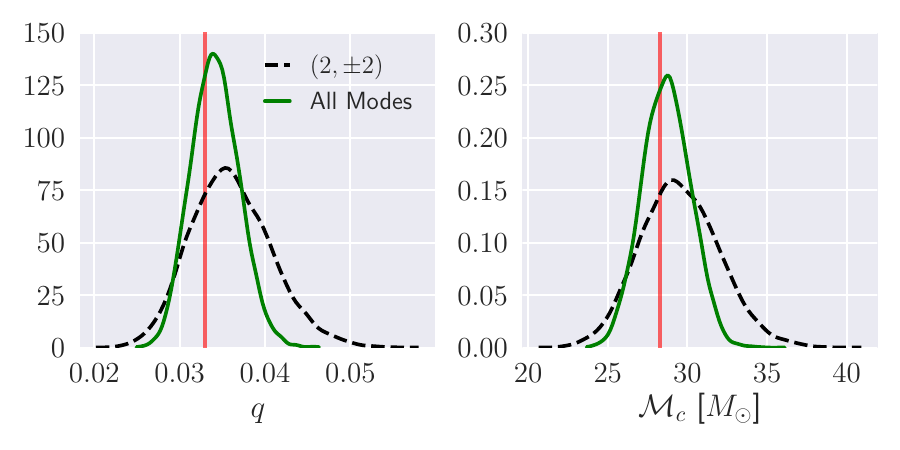}
		\caption{Mass posteriors recovered with (solid green) and without the higher modes (dashed black) for the binary with $M=225M_{\odot}$ and mass ratio $q=1/30$ in a face-off configuration. 
			Vertical lines show the true values.
			All other details are same as in Fig.\ref{Fig:chi2}.}
		\label{Fig:faceoff}
	\end{figure}

	\begin{figure}[hbt]
		\includegraphics[width=\columnwidth]{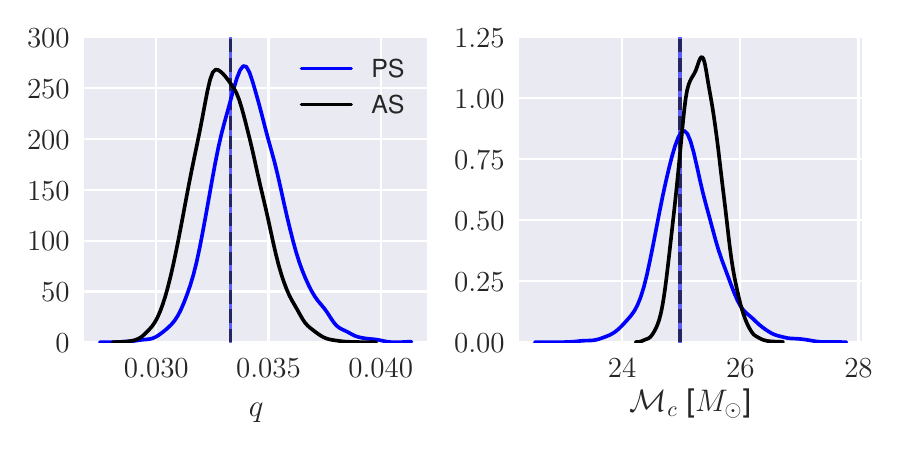}
		\caption{Same as in Fig.\ref{Fig:faceoff} but for binaries with $M=225M_{\odot}$, $q=1/30$ in aligned spin (AS) and generically precessing spins (PS) configuration. 
			Vertical lines show the true values.}
		\label{Fig:genericspin}
	\end{figure}
	
	\begin{figure}[bht]
		\includegraphics[width=\columnwidth]{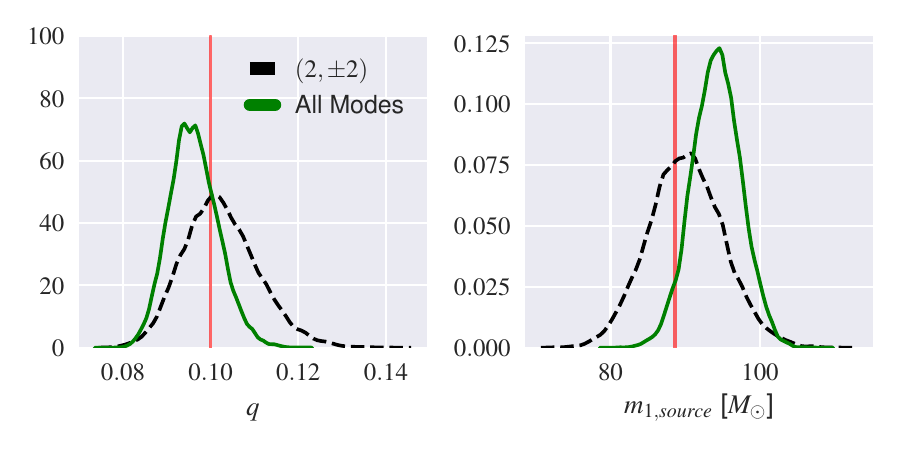}
		\caption{Mass posteriors recovered with (solid green) and without the higher modes (dashed black) for the binary with $M=120M_{\odot}$ and mass ratio $q=1/10$. 
			The source mass of the primary black hole lie in the pair-instability mass gap of $\sim50M_{\odot}<M<130M_{\odot}$. 
			Vertical lines show the true values.
			All other details are same as in Fig.\ref{Fig:chi2}.}
		\label{Fig:massgap}
	\end{figure}

	\paragraph{Face-off binaries: }
	Nearly all gravitational-wave observations to date have been characterized by 
	face-on (i.e. inclination angle $\iota=0.0$) or face-off (i.e. inclination angle $\iota=\pi$) orientation to the line-of-sight~\cite{2020arXiv201014527A}, minimizing the possibility of detecting higher order modes in general. 
	For such binaries, most of the higher order modes are expected to be weak. 
	However, we show that, for massive high mass-ratio binaries, higher order modes may have sufficient SNRs such that they can no longer be disregarded even for face-on/face-off cases. 
	To investigate that, we simulate a signal with total mass $M=225M_{\odot}$, mass ratio $q=1/30$, spins $\{\chi_1,\chi_2\}=\{0.6,0.7\}$ in face-off orientation such that the SNR is 25. 
	We then estimate the parameters using only $(2,\pm2)$ mode and with higher modes. 
	In Fig.\ref{Fig:faceoff}, we show the recovered posteriors with (solid green) and without (dashed black) the higher order modes. 
	We find that including the higher order modes help constrain the parameters better even for a face-off binary.
	
	\paragraph{Generic Spin Case: }
	While our work has exclusively focused on aligned-spin system, we now provide preliminary results for generic spin binaries.
	In Fig.\ref{Fig:genericspin}, we show two representative cases: we simulate signals with $\{M,q,\chi_1,\chi_2\}=\{225M_{\odot},1/30,0.6,0.7\}$ with aligned-spin and precessing spin configurations.
	For the generic spin case, we choose the spin angles as: $\theta_{1}=1.05$ and $\theta_{2}=1.02$, $\phi_{12}=3.53$ and $\phi_{jl}=3.75$ respectively (cf. Appendix of \cite{Romero-Shaw:2020owr}). 
	The signal is then recovered with generic spin \texttt{IMRPhenomXPHM} model. 
	We show the chirp mass and mass ratio posteriors recovered for both the cases. 
	We recover the mass source properties with similar accuracy. Future work should include a more 
	comprehensive investigation of generic spin IMRI systems.
	
	\paragraph{Mass-gap binaries: }
	\label{sec:mass-gap}
	Pair-instability and pulsational-pair-instability supernovae \cite{Woosley:2016hmi} prevents the formation of black holes with masses more than $\sim50M_{\odot}$ from stellar collapse. 
	This leads to a gap in the black-hole mass distribution function in between $\sim50M_{\odot}$ and $\sim130M_{\odot}$ \cite{Belczynski:2016jno, Renzo:2020rzx, Farmer:2019jed, Stevenson:2019rcw}. 
	The edges of the mass-gap region varies depending on the details of the pair-instability process, evolution of massive stars and core-collapse supernova explosion \cite{Mapelli:2017hqk, Mapelli:2019ipt}. 
	However, multiple stellar mergers and merger of black holes can lead to the formation of a black hole in the pair stability mass-gap region \cite{DiCarlo:2019fcq}. 
	It is therefore an interesting question to ask whether our results are valid for binaries with at-least one black hole that falls in the pair instability mass-gap. 
	We simulate a signal with $\{M,q\}=\{120M_{\odot},1/10\}$ with an SNR of 25. 
	The mass of the primary black hole is $109.1M_{\odot}$ ($88.35M_{\odot}$ in the source frame). 
	All other parameters are fixed to the default values used in this paper. 
	In Fig.\ref{Fig:massgap}, we show the recovered mass posteriors (in the source frame) with and without subdominant modes. 
	We find that 90\% credible regions are tighter when higher-order multiples are included in the recovery model.

	\section{Comparison with \texttt{SEOBNRv4HM\_ROM} results}
	\label{Sec:SEOB}
	
\begin{figure*}[bht]
		\includegraphics[scale=0.45]{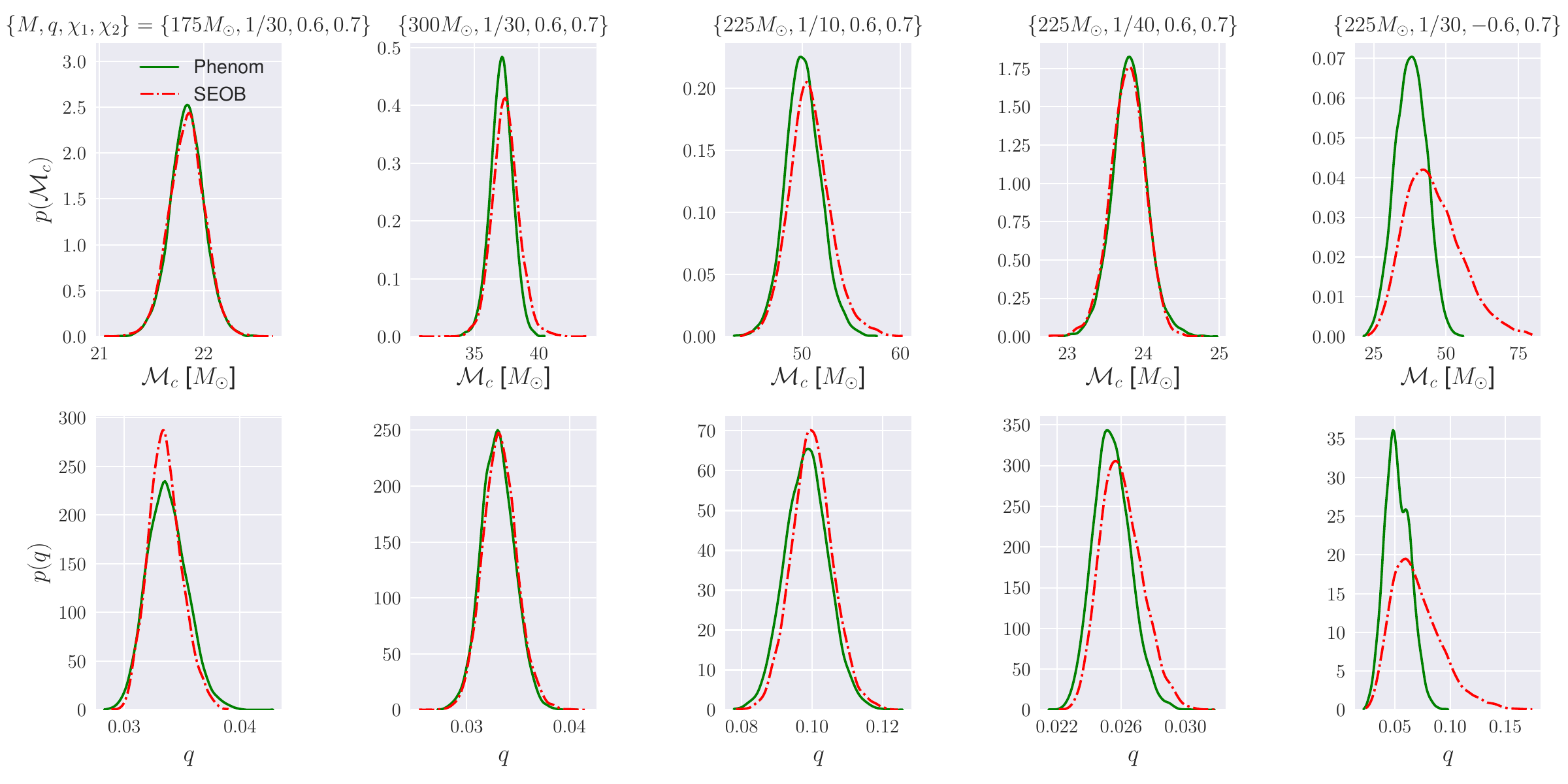}
		\caption{Chirp mass and mass ratio posteriors recovered with \texttt{IMRPhenomXHM} (solid green) and \texttt{SEOBNRv4HM\_ROM} (dash dotted red) for the binaries in the boundary of our parameter space.}
		\label{Fig:compare_Phenom_SEOB}
	\end{figure*}

	Our parameter estimation results are obtained using \texttt{IMRPhenomXHM} \cite{Garcia-Quiros:2020qpx}, a frequency-domain phenomenological waveform model which is calibrated to numerical relativity waveforms in the comparable mass ratio regime ($q>=1/18$) and to waveforms obtained from solving the perturbative Teukolsky equation for $1/1000>=q>=1/200$. As the model is uncalibrated in parts of the mass ratio regime we are looking at (i.e. $1/40<=q<=1/10$), we decide to redo the parameter estimation with a different waveform model for some of the representative cases. We choose the boundary cases in our parameter space (as listed in Table \ref{tab:1}) and employ \texttt{SEOBNRv4HM\_ROM} \cite{Cotesta:2020qhw}, an reduced order based effective-one-body model, in both injection and recovery. We find that the measurement uncertainties in different parameters do not change significantly. As an example, in Fig. \ref{Fig:compare_Phenom_SEOB}, we show the recovered posteriors for the chirp mass $M_c$ and mass ratio $q$ obtained using both \texttt{IMRPhenomXHM} and \texttt{SEOBNRv4HM\_ROM}.
	As none of the models are calibrated to NR simulations, it is not possible to prefer the results obtained using one of the approximants over others. Further, systematic analysis involving different waveform models in injection and recovery is also beyond the scope of the current paper. However, we expect the general parameter estimation trends presented in the paper, which have been compiled with \texttt{IMRPhenomXHM}, to be applicable to other IMRI waveform models.
	
	\section{Conclusion}
	\label{Sec:conlcusion}
	
	High mass-ratio massive binaries consisting of a stellar mass black hole and an IMBH, typically known as IMRIs,
	are a highly anticipated source of GWs for Advanced LIGO/Virgo. Due to the large total mass characterized by these systems,
	the number of in-band inspiral cycles (and therefore the power) from the dominant quadrapole mode is greatly reduced as 
	compared to comparable mass binaries. In this paper, by focusing on aligned spin systems with 
	detector-frame total masses $175 M_{\odot} < M < 300 M_{\odot}$ and mass ratios $1/40<q<1/10$, we show that 
	by including higher order harmonics into the analysis (i) results in a 3 to 4 times improvement in the measurement uncertainties, (ii) 10 times improvement in the recovered 3d comoving volume that contains the true position of the binary, (iii) constrain the spin magnitude of the primary and secondary black hole better than previously expected, and (iv) may improve the detectability of such binaries significantly. 
	We further show that current generation detectors are able to estimate the source properties of such binaries with $~1-15\%$ accuracy if higher order modes are included in the waveform model. 
	Omitting higher order modes, on the other hand, results in catastrophic parameter bias for many binary systems. 
	As one particularly striking example, a binary black hole system a with mass ratio of $q=1/30$ and whose  non-spinning primary BH has a mass of $m_1=218M_{\odot}$ is misclassified as a $q\sim1/4$ system with $m_1\sim162 M_{\odot}$ and $\chi_{1}\sim-0.88$ (Fig.\ref{Fig:90CI}). 
	These large parameter biases and measurement uncertainties would greatly reduces the science that can be extracted from IMRI signals, such as measuring the pair-instability mass-gap \cite{Belczynski:2016jno, Renzo:2020rzx, Farmer:2019jed, Stevenson:2019rcw}, distinguishing between IMRI formation channels \cite{AmaroSeoane:2007aw, Berry:2019wgg}, and self-consistency tests of GR that will be especially informative given the unique IMRI signal \cite{Gair:2012nm, Piovano:2020ooe,Yunes:2009ry, Canizares:2012ji, Canizares:2012is,Rodriguez:2011aa, Chua:2018yng}.
	Section \ref{Sec:SEOB} discusses the robustness of our results with respect to 
	waveform systematics, face-on binaries (which suppress higher-order modes)
	generic orbits with mis-aligned spins, and special cases such as when
	one of the component black holes lies in the pair-instability mass-gap ($\sim50M_{\odot}<m_{1}<130M_{\odot}$) \cite{Belczynski:2016jno, Renzo:2020rzx, Farmer:2019jed, Stevenson:2019rcw}.

	\prlsec{Acknowledgments}
	We thank Christopher Berry, Juan Calderon Bustillo, Harald Pfeiffer, Geraint Pratten, and Gaurav Khanna for helpful discussions and feedback.
	TI is supported by NSF grants PHY-1806665 and DMS-1912716, and a Doctoral Fellowship provided by UMassD Graduate Studies.
	SEF is partially supported by NSF grants PHY-1806665 and DMS-1912716.
	C.-J.H. acknowledge support of the National Science Foundation, and the LIGO Laboratory. 
	LIGO was constructed by the California Institute of Technology and Massachusetts Institute of Technology with funding from the National Science Foundation and operates under cooperative agreement PHY-1764464.
	The computational work of this project was performed on the CARNiE cluster at UMassD, which is supported by the ONR/DURIP Grant No.\ N00014181255. 
	A portion of this material is based upon work supported by the National Science Foundation under Grant No. DMS-1439786 while the author was in residence at the Institute for Computational and Experimental Research in Mathematics in Providence, RI, during the Advances in Computational Relativity program.
	This is LIGO Document Number DCC-P2100151.
	\bibliography{References}

\end{document}

%% file: preamble.tex
\usepackage[T1]{fontenc}
\usepackage[utf8]{inputenc}
\usepackage{lmodern}

\usepackage{verbatim}

\usepackage[dvipsnames, usenames]{xcolor}
\definecolor{linkcolor}{rgb}{0.0,0.3,0.5}
\usepackage[hypertexnames=false, unicode, colorlinks=true, linkcolor=linkcolor,
citecolor=linkcolor, filecolor=linkcolor,urlcolor=linkcolor,
pdfusetitle]{hyperref}

\usepackage[all]{hypcap}
\usepackage{graphicx}
\usepackage{xspace}
\usepackage{amssymb}
\usepackage{amsmath,amssymb,graphicx}
\usepackage{amssymb,amsmath}
\usepackage[normalem]{ulem}
\usepackage{bm}

\usepackage{microtype}

\usepackage[english]{babel}
\usepackage{blindtext}

\newcommand\prlsec[1]{\vspace{2mm}\noindent \textbf{\emph{#1}}.---}

\graphicspath{%
	{figs/}%
}

\DeclareMathAlphabet{\mathpzc}{OT1}{pzc}{m}{it}

\newcommand{\blambda}{\bm{\lambda}}
\newcommand{\btheta}{\bm{\theta}}

\newcommand{\n}{\mathbf{n}}

\usepackage{breqn}

\newcommand{\TITLE}{High-precision source characterization of intermediate mass-ratio black hole coalescences with gravitational waves: The importance of higher-order multipoles}

\newcommand{\LIGOlabMIT}{\affiliation{LIGO Laboratory, Massachusetts Institute of Technology, 185 Albany St, Cambridge, MA 02139, USA}}
\newcommand{\MKI}{\affiliation{Department of Physics and Kavli Institute for Astrophysics and Space Research, Massachusetts Institute of Technology, 77 Massachusetts Ave, Cambridge, MA 02139, USA}}
\newcommand{\monash}{\affiliation{School of Physics and Astronomy, Monash University, Vic 3800, Australia}}
\newcommand{\ozgrav}{\affiliation{OzGrav: The ARC Centre of Excellence for Gravitational Wave Discovery, Clayton VIC 3800, Australia}}
\newcommand{\UMassDMath}{\affiliation{Department of Mathematics,
		University of Massachusetts, Dartmouth, MA 02747, USA}} 
\newcommand{\UMassDPhy}{\affiliation{Department of Physics,
		University of Massachusetts, Dartmouth, MA 02747, USA}} 
\newcommand{\CSCVR}{\affiliation{Center for Scientific Computing and Visualization Research,
		University of Massachusetts, Dartmouth, MA 02747, USA}} 
\newcommand{\KITP}{\affiliation{Kavli Institute for Theoretical Physics, University of California, Santa Barbara, CA 93106, USA}}

%% file: authors.tex
\author{Tousif Islam}
\email{tislam@umassd.edu}
\UMassDPhy
\UMassDMath
\CSCVR
\KITP
\author{Scott E. Field}
\UMassDMath
\CSCVR
\author{Carl-Johan Haster}
\LIGOlabMIT
\MKI
\author{Rory Smith}
\monash
\ozgrav

\hypersetup{pdfauthor={Islam et al.}}